\documentclass{article}
\usepackage{spconf}
\usepackage{graphicx}
\usepackage[noadjust]{cite}
\usepackage{amsmath,amssymb,amsfonts}
\usepackage{algorithm}
\usepackage{algpseudocode}
\usepackage{tabularx}
\usepackage{booktabs}
\usepackage{graphicx}
\usepackage{textcomp}
\usepackage{graphicx}
\usepackage{url}
\usepackage{multirow} 
\usepackage{rotating}
\usepackage{subcaption} 
\usepackage{flushend}

\newcommand{\x}{\mathbf{x}}
\newcommand{\y}{\mathbf{y}}

\newcommand\blfootnote[1]{%
  \begingroup
  \renewcommand\thefootnote{}\footnote{#1}%
  \addtocounter{footnote}{-1}%
  \endgroup
}

\title{Denoising Diffusion Probabilistic Models for Magnetic Resonance Fingerprinting}
\begin{document}
\name{%
\begin{tabular}{@{}c@{}}
Perla Mayo \quad 
Carolin M. Pirkl \quad 
Alin Achim \quad
Bjoern H. Menze \quad
Mohammad Golbabaee
\end{tabular}
}
\address{}

\maketitle
\begin{abstract}
Magnetic Resonance Fingerprinting (MRF) is a time-efficient approach to quantitative MRI, enabling the mapping of multiple tissue properties from a single, accelerated scan. However, achieving accurate reconstructions remains challenging, particularly in highly accelerated and undersampled acquisitions, which are crucial for reducing scan times. While deep learning techniques have advanced image reconstruction, the recent introduction of diffusion models offers new possibilities for imaging tasks, though their application in the medical field is still emerging. Notably, diffusion models have not yet been explored for the MRF problem. In this work, we propose for the first time a conditional diffusion probabilistic model for MRF image reconstruction. Qualitative and quantitative comparisons on in-vivo brain scan data demonstrate that the proposed approach can outperform established deep learning and compressed sensing algorithms for MRF reconstruction. Extensive ablation studies also explore strategies to improve computational efficiency of our approach.

\end{abstract}

\blfootnote{Firstly submitted on October, 2024. PM, AA and MG are with the University of Bristol, UK. CMP is with GE HealthCare, Munich, Germany. BHM is with the University of Zurich, Switzerland. PM and MG are the corresponding authors (e-mail: pm15334@bristol.ac.uk and m.golbabaee@bristol.ac.uk, respectively).}

\section{Introduction}
\label{sec:intro}
Quantifying the intrinsic nuclear magnetic resonance (NMR) properties of tissues~\cite{tofts2005quantitative}, such as T1 and T2 relaxation times, provides a powerful tool for tissue characterization and monitoring of pathological changes. However, conventional quantitative MRI (qMRI) techniques face challenges due to long acquisition times, limiting their clinical utility. This is because qMRI often requires sampling a long  sequence (time series) of magnetization images to capture a single tissue property per scan, making it difficult to acquire the full spatiotemporal volume required for encoding the parameter information of interest in a reasonable scan time. As encoding-efficient alternatives to conventional steady-state parameter mapping techniques, Magnetic Resonance Fingerprinting (MRF)~\cite{ma2013mrf} and other transient-state imaging techniques~\cite{jiang2015fisp,gomez2019qtimrf,cao2022tgas} were introduced to address this problem. To reduce scan time, MRF uses short-length acquisition sequences to encode multiple tissue properties simultaneously and applies compressed sensing to subsample only a fraction of the spatiotemporal k-space data. However, faster scans lead to challenges in image reconstruction, including aliasing artifacts from k-space undersampling and limited tissue property information due to the truncated acquisition sequences. Effective image reconstruction algorithms are needed to tackle these challenges and improve the accuracy and precision of tissue parameter estimation.

\subsubsection*{MRF Imaging Algorithms} 
Early MRF works~\cite{ma2013mrf,mcgivney2014svdmrf} used gridding reconstruction (non-Cartesian Fourier back-projection) to convert undersampled k-space data into image time series. Tissue properties were then estimated through dictionary matching, where voxel time signals were compared to a precomputed dictionary of theoretical tissue fingerprints from Bloch equations. However, the accuracy of these methods was limited by aliasing artifacts that were not addressed during the image reconstruction step. Later compressed sensing models used iterative dictionary matching~\cite{asslander2018admm,davies2014blip,golbabaee2019coverblip} to enforce time-domain priors during reconstruction. Papers~\cite{asslander2018admm,zhao2018lowrank} showed that a low-rank subspace model driven by the Singular Value Decomposition (SVD) of MRF dictionary could efficiently act as a time-domain prior, while reducing significant computation overhead. Subsequent compressed sensing approaches~\cite{mazor2018flor,hamilton2019simultaneous,lima2019sparsity,hu2021highqualitymrf} additionally incorporated some form of sparsity and/or low rank spatial priors to reduce undersampling artefacts. However, the most effective MRF algorithms to date are data driven, based on deep learning. The initial MRF deep learning works~\cite{hoppe2017dl4mrf,cohen2018drone,golbabaee2019geometry} replaced the dictionary matching step with pixel-wise neural network models trained solely on the MRF dictionary, without incorporating anatomical information. Consequently, these models had limited capacity to handle spatially correlated aliasing artifacts. More recent deep learning works~\cite{balsiger2018mrf,fang2019supervisedmrf,soyak2021convica,fatania2022plug,hamilton2022dipmrf,li2023learnedtensor} use anatomical MRF imaging datasets to train CNNs with various architectures and losses to learn more effective spatiotemporal priors directly from data.
Some models e.g.~\cite{fatania2022plug} aim to restore (de-alias) MRF image time series, while tissue quantification is performed by standard dictionary matching. Other models e.g.~\cite{fang2019supervisedmrf} combine the two steps and directly output restored quantitative maps. Reconstructing MRF image time series, as opposed to directly estimating parameter maps, could provide a more transparent pipeline, ensuring parameter maps adhere to Bloch equations and, by extension, the underlying physics of the model. 

\subsubsection*{Diffusion models} 
Denoising diffusion probabilistic models (DDPM) \cite{dpm2015,ho2020ddpm,nichol2021iddpm,song2021ddim} have recently gained significant attention due to their remarkable performance on image generation~\cite{dhariwal2021ddpmbeatgans,rombach2022hrimagesynth,hoogeboom2023simplediffusion} and image restoration tasks~\cite{choi2021ilvr,xie2023smartbrush,saharia2022imagesriterative,ozdenizci2023patchddpm,xia2023diffir}. Built on a hierarchy of denoising autoencoders, DDPMs are trained to iteratively refine image details by denoising and reversing a diffusion process. By incorporating conditional information, DDPMs can be guided during sample image generation~\cite{nichol2021iddpm}, for example, in image restoration tasks, the model is conditioned on low-quality degraded image inputs, producing high-quality restored images as the outcome~\cite{ozdenizci2023patchddpm,saharia2022imagesriterative}. 
While DDPM applications in medical imaging are emerging (e.g. in segmentation~\cite{wu2024medsegdiff,wolleb2022diffusionmedseg}, image-to-image translation~\cite{ozbey2023adversarialdiffusion,li2023zeroshottranslation}, and image reconstruction~\cite{liu2023dolce,xu2023zeroshotct,peng2022towardsmrdiffusion,bian2024qmri,wang2025qmri}), their potential for MRF is yet unexplored. Compared to CNNs, the current state-of-the-art for MRF reconstruction, DDPMs offer several advantages. DDPMs' iterative refinement allows for stage-wise, progressive restoration of image details from coarse to fine scales, leading to more precise recovery compared to the single-pass structure of CNNs. 
They also incorporate advanced mechanisms like self-attention~\cite{vaswani2017attention} or transformers~\cite{dosovitskiy2020transformers}, which more effectively model long-range pixel dependencies. Additionally, while CNNs treat image restoration as a deterministic mapping from a corrupted to a restored image, DDPMs operate within a probabilistic framework, allowing them to learn the distribution of possible solutions for a given task and generate multiple samples i.e. a useful feature for capturing uncertainties in the reconstructed images.

\subsubsection*{Our work}
This paper presents, to the best of our knowledge, the first diffusion-based model for MRF reconstruction to enable efficient multi-parametric estimations of T1 and T2 relaxation maps from highly accelerated acquisitions. Our approach, coined as MRF-IDDPM, is trained as a conditional DDPM model to generate restored MRF image time series from initial gridding reconstructions, which suffer from significant aliasing artifacts due to accelerated MRF acquisitions. Evaluated on in-vivo brain scans data, we show that the proposed approach outperforms established deep learning and compressed sensing MRF algorithms in terms of image reconstruction and quantitative mapping accuracies. Additionally, the probabilistic nature of DDPM allows us to generate multiple sample reconstructions given an input and produce a variance map to visualize reconstruction uncertainties. 
To address the computational challenges associated with training diffusion models on high-dimensional MRF data, we employed a subspace dimensionality reduction approach from~\cite{mcgivney2014svdmrf} to temporally compress the MRF image time series, enabling the diffusion process to operate within a reduced latent space. We further implemented a patch-based processing pipeline to efficiently handle the spatial dimensions of data, enabling training and reconstruction on modest GPUs with limited memory and reasonable runtimes. 

The rest of the manuscript is structured as follows: in Section~\ref{sec:preliminaries} we provide the preliminaries for our method, which is then described in more details in Section~\ref{sec:the_algorithm}. Numerical evaluations and comparisons with baselines techniques are provided in Section~\ref{sec:experiments}, alongside the results of ablation studies and a discussion of limitations and future directions. Our conclusions are provided in Section~\ref{sec:conclusions}.

\section{Preliminaries}
\label{sec:preliminaries}

\subsection{The MRF Reconstruction Problem}
\label{subsec:prelim_mrf_problem}
The MRF reconstruction is an inverse problem:
\begin{equation}
    \label{eq:forward_operator}
    \textbf{y} \approx \underbar{A}(\mathbf{\underbar{x}}),\; 
    \textrm{s.t.}\; \mathbf{\underbar{x}}_v={\boldsymbol\rho}_v  \mathcal{B}(\text{T1}_v,\text{T2}_v), \;\forall v : \text{voxels}, 
\end{equation}
where, $\textbf{y} \in \mathbb{C}^{c \times m\times l}$ are the undersampled k-space measurements from $m$ spatial frequency locations across $c$ receiver coil channels, and $l$ time frames. The linear acquisition operator $\underbar{A}: \mathbb{C}^{n \times l} \rightarrow \mathbb{C}^{c \times m \times l}$ includes coil sensitivities and the undersampled spatial Fourier operations at each time frame. Given $\y$, the aim is to reconstruct the time series of magnetization images (TSMI) $\mathbf{\underbar{x}} \in \mathbb{C}^{n \times l}$ that contains the tissue-encoding signals of length $l$ for $n$ voxels. The TSMI and the desired tissue parameter maps (Q-Maps), here the spatial maps of T1 and T2 relaxations times, are voxel-wise related through the non-linear Bloch response model $\mathcal{B}: \mathbb{R}^2\rightarrow \mathbb{C}^l$, scaled by proton density $\boldsymbol\rho$. The MRF dictionary $\textbf{\underbar{D}}$ is a matrix that contains precomputed Bloch responses for a discretized fine grid of (T1, T2) values, and it acts as a lookup table linking tissue properties to corresponding magnetization time signals or fingerprints. Once the TSMI $\mathbf{\underbar{x}}$ is reconstructed, a voxel-wise dictionary matching process is applied to identify the T1 and T2 maps~\cite{ma2013mrf}. McGivney et al.~\cite{mcgivney2014svdmrf} proposed a low-rank SVD decomposition of the MRF dictionary, approximating $\textbf{\underbar{D}}\approx \textbf{D}\textbf{V}^H$, where $\textbf{D}$ is the dimensionally-reduced dictionary in an $s$-dimensional subspace ($s \ll l$) defined by the orthonormal basis $\textbf{V}\in \mathbb{C}^{l\times s}$. They also show that TSMI can be factorized in this basis as $\mathbf{\underbar{x}} \approx \mathbf{\underbar{x}}\textbf{VV}^{H} = \x\textbf{V}^{H}$, where $\x= \underbar{x}\textbf{V}$ represents the dimensionally-reduced TSMI. With this reduction, our reconstruction problem becomes estimating a dimension-reduced TSMI from $\y \approx A(\x)$, where $A = \underbar{A}\circ \textbf{V}^H$. Unless otherwise specified, this dimension reduction is consistently applied throughout our work.

The MRF acquisition process can be greatly shortened by combining k-space undersampling with a reduction in the number of measured time frames  
$l$. However this approach complicates the reconstruction problem by introducing significant aliasing artifacts, limited tissue parameter encoding and increased noise sensitivity --- challenges we aim to address in this work.

\subsection{Denoising Diffusion Probabilistic Models for Image Reconstruction}
\label{subsec:prelim_ddpm}
Denoising Diffusion Probabilistic Models (DDPM)~\cite{dpm2015,ho2020ddpm} describe a Markovian process in which a sample image from the data distribution $q(\x_0)$ is gradually corrupted by adding Gaussian noise over $t=1,\ldots, T$ diffusion steps, generating latent variables $\x_t$ with increasing noise levels $\beta_t \in (0, 1)$. For a sufficiently large $T$, the variable $\x_T$ approximates pure noise with an isotropic Gaussian distribution. The stepwise \emph{forward process} is defined as:

\begin{equation}
    q(\x_t |\x_{t-1}) = \mathcal{N}(\x_t;\sqrt{1-\beta_t}\x_{t-1}, \beta_t\textbf{Id}),
    \label{eq:q_normal}    
\end{equation}

where $\mathcal{N}(x;\mu,\sigma^2\textbf{Id})$ denotes  a Gaussian distribution with mean $\mu$ and diagonal covariance $\sigma^2\textbf{Id}$. Using cumulative noise schedules $\alpha_t=1-\beta_t$, $\bar{\alpha}_t=\prod_{j=0}^{t}\alpha_j$, $\x_t$ can be directly derived from the clean sample $\x_0$ as:

\begin{equation}
    \x_t = \sqrt{\bar{\alpha}_t}\x_0 + \sqrt{1-\bar{\alpha}_t}\epsilon, \,\,\,\,\text{where}\,\,\,\, \epsilon \sim \mathcal{N}(\textbf{0}, \textbf{Id}),
    \label{eq:sampling_xt}
\end{equation}

without having to progress through all previous diffusion steps. DDPMs train a neural network $\epsilon_\theta$ to estimate the noise $\epsilon$ at each diffusion step, with a learning objective to minimize the loss:

\begin{equation}
    L_{\epsilon} = \mathbb{E}_{\x_0, t, \epsilon}\left[\| \epsilon - \epsilon_{\theta}(\x_t, t)\|_2^2\right].
    \label{eq:loss_noise}
\end{equation}

In the \emph{reverse process}, starting from a noise vector $\x_T \sim \mathcal{N}(\textbf{0}, \textbf{Id})$, we iteratively compute: 

\begin{equation}
    \x_{t-1}=\frac{1}{\sqrt{\alpha_t}}\left(\x_t - \frac{1-\alpha_t}{\sqrt{1-\bar{\alpha}_t}}\epsilon_{\theta}(\x_t, t)\right)+\sigma_t \textbf{z},
\end{equation}

where $\textbf{z} \sim \mathcal{N}(\textbf{0}, \textbf{Id})$ when $t>1$ and $\textbf{0}$ otherwise, and the variances $\sigma_t$, as per~\cite{ho2020ddpm}, are set to some pre-defined values determined by $\beta_t$. During the reverse process, a sample is gradually de-noised from high to low noise levels, recovering coarse to fine image details progressively. After $T$ steps, this process generates an approximate sample from the distribution $q(\x_0)$.

\subsubsection{Faster Reverse Process}
\label{subsec:iddpm}
Practical diffusion steps $T$ are typically large. To accelerate the reverse process, Nichol and Dhariwal~\cite{nichol2021iddpm} proposed Improved Denoising Diffusion Probabilistic Models (IDDPM), which incorporate a variance-learning term into to the loss function:

\begin{equation}
    L = L_{\epsilon} + \lambda L_{\sigma},
    \label{eq:loss_hybrid}
\end{equation}

where $L_{\sigma}$ optimizes the learned variances $\sigma_t=\sigma_{\theta}(\x_t, t)$ as an additional output of the network (besides estimating noise), and $\lambda$ to balance the loss terms. Nichol and Dhariwal showed that learning the variances enables high-quality sample generation in far fewer $K \ll T$ steps. This efficient approach is adopted in our MRF reconstruction pipeline (detailed in Algorithm 2).

\subsubsection{Conditional DDPMs}
\label{subsec:conditional_iddpm}
The DDPM denoising network $\epsilon_\theta$ can be conditioned on auxiliary inputs, enabling it to generate samples constrained by specific prior information. In the context of image restoration~\cite{saharia2022imagesriterative, ozdenizci2023patchddpm}, a degraded image $\x_c$ can be concatenated with the noisy latent variable $\x_t$ along the channel dimension, to guide the model through the diffusion process. The network is thus parametrised as $\epsilon_{\theta}(\x_t, t, \x_c)$ so to refine its denoising function by adapting it to the given conditioning input. This setup allows the diffusion model to learn a structured mapping between the degraded input and the target conditional distribution $q(\x_0 | \x_c)$ of clean images. 
\section{The MRF-IDDPM Algorithm}
\label{sec:the_algorithm}
We are now ready to describe our conditional DDPM approach for generating restored MRF image time series (TSMI) samples from highly accelerated k-space data. The pipeline comprises three main stages: preprocessing, training, and inference, each detailed below.

\subsection{Preprocessing}
\label{subsec:algorithm_preprocessing}
In the preprocessing step, we start with k-space measurements $\y_{\text{ref}}$ from extended reference MRF acquisitions such as~\cite{jiang2015fisp}. These can be used to generate low-aliasing quantitative maps via compressed sensing reconstruction methods like LRTV~\cite{golbabaee2021lrtv}. From these quantitative maps we then produce high-quality (low-aliasing) reference TSMIs $\x_{\text{ref}}$ using the Bloch response model in Eq.~\eqref{eq:forward_operator}. Next, we subsample $\y_{\text{ref}}$ by truncating it along the time dimension to obtain k-space data $\y$, representing retrospectively faster acquisitions. Reconstruction of these accelerated acquisitions using a basic gridding algorithm $\x_{\text{grid}}=A^H\y$, or a compressed sensing method leads to severe aliasing artifacts.

The dataset for training our diffusion model consists of pairs $\{\x_c=\x_{\text{grid}}, \x_0=\x_{\text{ref}}\}$ obtained from multiple MRF scans. For computational efficiency, both $\x_c$ and $\x_{\text{ref}}$ undergo dimensionality reduction using the SVD basis (with $s=5$ in our experiments) of the MRF dictionary~\cite{mcgivney2014svdmrf}. Our base model assumes that $\x_c$ and $\x_{\text{ref}}$ have the same (truncated) time length, allowing the use of a shared SVD basis for compression. However, we also experimented with a model that outputs $\x_0$ in an untruncated time length, matching the longer reference acquisition sequence. In this case, the SVD bases for $\x_c$ and $\x_{\text{ref}}$ differ, as the latter is derived from the decomposition of a dictionary with longer signals/fingerprints. To process the complex-valued $\x_c$ and $\x_0$, we concatenated the real and imaginary parts along the channel dimension. Pixel values in both $\x_c$ and $\x_{\text{ref}}$ were divided by two fixed constants, respectively, to normalize their values within the range [-1, 1] across the entire training dataset. These values were determined by inspecting the histograms of TSMI pixel intensities in our dataset.

\subsection{Training}
\label{subsec:training}
MRF data is high-dimensional, making diffusion process training computationally challenging. While SVD-based dimensionality reduction helps manage TSMIs' temporal dimension, we address the spatial dimension complexity by implementing a patch-wise training and inference pipeline. As shown in our experiments, this patch-based approach allows MRF data to fit on modest GPU memory and significantly accelerates training. Additionally, patch-level training enables greater data augmentation opportunities, especially for small MRF image datasets. For sufficiently large patches that capture essential spatial context, we show that the model can still produce high-quality MRF reconstructions.

The training process is outlined in Algorithm~\ref{alg:training}. We choose a denoising model $\epsilon_\theta$ from~\cite{nichol2021iddpm}. The model is trained on TSMI patches $\{\x_{c}^{(i)} = \text{Crop}(\textbf{P}_i, \x_c)$, $\x_{0}^{(i)} = \text{Crop}(\textbf{P}_i, \x_0)\}$ of dimension $p \times p$ extracted using binary masks $\textbf{P}_i$ at random locations $i$ from the full size pre-processed images $\x_{c}$ and $\x_0$, respectively. Consequently, the latent noisy data $\x_t^{(i)}$ is also treated as a patch. The low-quality TSMI patches $\x_c^{(i)}$ are channel-concatenated to $\x_t^{(i)}$ as inputs of the denoising network $\epsilon_\theta$ to guide the conditional diffusion process. To improve generalization, data augmentation was employed with random vertical/horizontal spatial flips of image patches at each training iteration. We use $T=1000$ diffusion steps with an evenly spaced linear noise schedule $\beta_t$ where $\beta_0 = 0.0001$ and $\beta_T = 0.02$. Our base model uses patch size $p=64$, which we experimentally show to be a good compromise between accuracy and computational time. Training optimizes the loss function in~\eqref{eq:loss_hybrid} using ADAM optimiser for 150K iterations with fixed learning rate $10^{-4}$, effective batch size 32, dropout rate $0.3$, and an exponential moving average scheme with exponential rate 0.9999 to average model parameters for stable training~\cite{nichol2021iddpm}.

\label{subsec:algorithm_training}
\begin{algorithm}[t]
    \caption{MRF-IDDPM Training}\label{alg:training}
    \begin{algorithmic}[1]
        \Require Clean and aliased MRF images $(\x_0, \x_c)$, diffusion steps $T$, cumulative noise schedules $\bar{\alpha}_t$. 
        \Repeat
            \State Sample a random binary mask $\textbf{P}_i$
            \State $\x_0^{(i)} = \text{Crop}(\textbf{P}_i, \x_0)$, $\x_c^{(i)} = \text{Crop}(\textbf{P}_i, \x_c)$
            \State $t\sim \text{Uniform}(\{1\hdots T\})$
            \State $\epsilon \sim \mathcal{N}(\textbf{0}, \textbf{Id})$
            \State Take a gradient step on
            \State \hspace*{5mm} $\nabla_{\theta}||\epsilon -\epsilon_{\theta}(\sqrt{\bar{\alpha}_t}\x_0^{(i)} + \sqrt{1-\bar{\alpha}_t}\epsilon, t, \x_c^{(i)})||^2 + \lambda L_{\sigma}$ 
        \Until convergence \newline
        \Return $\epsilon_\theta$
    \end{algorithmic}
\end{algorithm}

\subsection{Inference by Reverse Diffusion}
\label{subsec:algorithm_sampling}
Algorithm~\ref{alg:inference} describes the iterative reverse diffusion process for MRF reconstruction so to obtain restored TSMI samples $\hat{\x}_0$ from an initial gridding reconstruction as the conditioning image $\x_c$. We divide the TSMI image into overlapping spatial patches. Starting with each patch $\x_T^{(i)}$ initialized as pure noise, the network $\epsilon_{\theta}$ performs reverse iterations, where it takes the conditioned patch $\x_c^{(i)}$ alongside the noisy patch $\x_t^{(i)}$ to denoise the latter and estimate $\x_{t-1}^{(i)}$. Similar to training, each conditional image is range-normalized to [-1, 1]. Our strategy for patch inference consists of running the complete reverse process on each patch, using IDDPM as a fast sampling backbone, which allows for reconstruction in as few as $K=50\ll T$ steps. After retrieving all overlapping patches of the restored image, they are aggregated and averaged in the common regions (lines 11 to 14). As shown in our experiments, careful selection of patch size and stride during the reverse process is essential for balancing accuracy and computational efficiency. Finally, the reconstructed full-size TSMI, $\hat{\x}_0$, is processed by dictionary matching to obtain the T1 and T2 parameter maps. Repeating Algorithm~\ref{alg:inference} with different random draws for the initial noise $\x_T$ enables sampling variation across generated reconstructions, for a given condition image $\x_c$.

\begin{algorithm}[t]
    \caption{MRF-IDDPM Inference}\label{alg:inference}
    \begin{algorithmic}[1]
        \Require Aliased MRF image $\x_c$, denoising diffusion model $\epsilon_{\theta}(\x_t, \x_c, t)$, sampling steps $K$, patch locations $D$.
        \State $\textbf{M}=\textbf{0}$
        \State $\x_T \sim \mathcal{N}(\textbf{0}, \textbf{Id})$
        \For{$i \in D$}
            \State $\x_c^{(i)} = \text{Crop}(\textbf{P}_i, \x_c), \x_T^{(i)} = \text{Crop}(\textbf{P}_i, \x_T)$
            \For{$\text{k}=K,\hdots, 1$}
                \State $t=(k-1)\cdot T/K + 1$
                \State $t_{\text{next}}=(k-2)\cdot T/K + 1$ if $k>1$ else $0$
                \State $\textbf{z} \sim \mathcal{N}(\textbf{0}, \textbf{Id})$ if $k>1$ else $\textbf{0}$
                \State $\x_{t_{\text{next}}}^{(i)}=\frac{1}{\sqrt{\alpha_t}}\left(\x_t^{(i)} - \frac{1-\alpha_t}{\sqrt{1-\bar{\alpha}_t}}\epsilon_{\theta}(\x_t^{(i)}, t, \x_c^{(i)})\right)+{\boldsymbol\sigma}_t \textbf{z}$
            \EndFor
            \State $\hat{\x}_0 = \hat{\x}_0 + \text{Uncrop}(\textbf{P}_i, \x_{t_{\text{next}}}^{(i)})$
            \State $\textbf{M}= \textbf{M} + \textbf{P}_i$
        \EndFor
        \State $\hat{\x}_0 = \hat{\x}_0 ./ \textbf{M}$ \newline
        \Return $\x_0$
    \end{algorithmic}
\end{algorithm}
\section{Numerical Experiments}
\label{sec:experiments}
\subsection{Experimental Setup}
\label{subsec:exps_setup}
The proposed method was evaluated along with the baseline approaches on 2D healthy volunteer brain MRF data, acquired using the Steady State Precession (FISP) with the same flip angle schedule as in~\cite{jiang2015fisp} with repetition time, echo time and inversion time (TR/TE/TI) of 10/1.908/18 ms, respectively, $l=1000$ repetitions (time frames), non-Cartesian k-space sampling with variable density spiral readouts, matrix size of $n=230\times230$ with 1mm in-plane resolution and 5mm slice thickness. Data was acquired on a 3T MRI scanner (MR750w system GE HealthCare, Waukesha, WI) with an 8-channel receive-only head RF coil. \footnote{This work was conducted under institutional approval for EPSRC grant EP/X001091/1. The retrospective study used anonymized human scans provided by GE Healthcare, with informed consent obtained in compliance with the German Medical Devices Act.}  
Our described pipeline was employed for the reconstruction of $R=5$ fold accelerated scans by retrospective truncation of the FISP sequence length to $l=200$. A total of 8 subjects, 15 axial slices each were available. From this, we use 6 subjects for training and the rest for evaluation/testing. The reference Q-Maps were reconstructed using the baseline LRTV method~\cite{golbabaee2021lrtv} from full-length ($l=1000$) FISP acquisitions. To evaluate the performance of TSMI reconstructions, we employed references estimated from the clean Q-Maps following the Bloch response model defined in Eq.~\eqref{eq:forward_operator}. All experiments used a small NVIDIA 2080Ti GPU for computations, and for the training and testing of all deep learning pipelines.

\begin{table*}[hbt!]
    \centering
    \footnotesize 
    \begin{tabular}{l|cc|cc|ccc|ccc}
    \hline
    \hline
         \multirow{ 2}{*}{\textbf{Method}} & \multicolumn{2}{ c|}{\textbf{MAPE (\%)} $\downarrow$} & \multicolumn{2}{ c|}{\textbf{RMSE (ms) $\downarrow$}} & \multicolumn{3}{ c|}{\textbf{NRMSE} $\downarrow$}& \multicolumn{3}{ c}{\textbf{SSIM} $\uparrow$} \\ \cline{2-11}
                & \textbf{T1} & \textbf{T2} & \textbf{T1} & \textbf{T2} & \textbf{T1} & \textbf{T2} & \textbf{TSMI} & \textbf{T1} & \textbf{T2} & \textbf{TSMI} \\ 
                \hline 
         \textbf{SVDMRF}   & 16.98 & 140.86 & 79.39 & 36.07 & 0.5566 & 1.8710 & 3.8360 & 0.9221 & 0.7547 & 0.7219 \\
         \textbf{LRTV}     & 15.94 &  35.96 & 76.62 & 15.40 & 0.5349 & 0.7353 & 1.2377 & 0.9376 & 0.8898 & 0.7932 \\
         \textbf{SCQ}       &  9.58 & 23.67 & 16.22 &  9.16 & 0.1143 & 0.4354 &  --    & 0.9750 & 0.9674 & -- \\ 
         \textbf{DRUNet}    &  8.93 & 21.07 & 18.28 &  9.39 & 0.1285 & 0.4458 & 0.5633 & 0.9693 & 0.9588 & 0.8388 \\ 
         \textbf{MRF-IDDPM (Ours)} & \textbf{7.19} & \textbf{17.64} & \textbf{14.10} & \textbf{7.63} & \textbf{0.1000} & \textbf{0.3669} & \textbf{0.4142} & \textbf{0.9759} & \textbf{0.9712} & \textbf{0.8729} \\ 
         \hline
         \hline 
    \end{tabular}
    \caption{Reconstruction metrics (averaged over test dataset) for TSMIs and T1/T2 parameter maps  using evaluated algorithms.}
    \label{tab:metrics_selected_slices}
\end{table*}

\subsection{Tested Algorithms}
\label{subsec:exps_benchmarks}
We compare our approach with classical compressed sensing and deep learning methods for MRF reconstruction. Tested algorithms, except SCQ, used dictionary matching to map the reconstructed TSMIs to T1 and T2 maps. For this purpose, an MRF dictionary consisting of 95K fingerprints was simulated on a logarithmically-spaced grid of $(T1,T2) \in [0.01,6] \text{s} \times[0.004,4] \text{s}$) using the extended-phase-graph formalism~\cite{weigel2015epg}. Tested methods also used SVD dimensionality reduction with $s=5$ to reconstruct TSMIs (see \ref{subsec:prelim_mrf_problem}), except for SCQ which uses a different dimensionality reduction learned by a neural network. Below we describe each tested baselines:

\subsubsection{MRF-IDDPM}
\label{subsub:exps_mrfiddpm}
Our implementations used the publicly available code for IDDPM\footnote{\url{https://github.com/openai/improved-diffusion}} in PyTorch. The architecture follows the U-Net structure described in~\cite{nichol2021iddpm}, consisting of four downsampling/upsampling blocks (starting with 128 channels), two residual blocks, and one attention head introduced at a resolution of $16 \times 16$. During inference, $K = 50$ IDDPM sampling steps were used instead of the $T = 1000$ diffusion steps for training. Our base model reconstruct the TSMI from overlapping patches of size 64 and stride 8. To evaluate performance, we run the reconstruction process ten times with different random initializations of $\x_T$, and averaged the TSMI results before dictionary matching. Our ablation studies (section \ref{subsec:exps_ablation}) demonstrate that these settings are sufficient to reconstruct high-quality tissue parameter maps.

\subsubsection{SVDMRF}
A non-iterative reconstruction technique based on gridding $\x_{\text{grid}}=A^H\y$, as proposed in \cite{mcgivney2014svdmrf}. The TSMIs obtained by this method are the same as the conditioning images used in our MRF-IDDPM pipeline for further processing and image enhancement.

\subsubsection{LRTV} A compressed sensing method~\cite{golbabaee2021lrtv} based on convex optimisation to reconstruct TSMIs from undersampled k-space measurements by enforcing k-space data consistency and applying spatial Total Variation regularization.

\subsubsection{TSMI restoration CNN}
\label{subsubsec:dl4tsmienhancing}
We employed Zhang's Deep Residual U-Net (DRUNet) model~\cite{zhang2021drunet}, a state-of-the-art CNN recently used for MRF reconstruction in~\cite{fatania2022plug,fatania2023equivariant}. We trained the network for TSMI restoration task; similar to our MRF-IDDPM, the inputs are aliasing-contaminated TSMIs from gridding reconstruction of faster MRF acquisitions, and outputs are the reference TSMIs from longer acquisitions. Similar to~\cite{fatania2023equivariant}, the residual U-Net architecture had four down/upsampling blocks staring with 32 feature channels, ReLU activations, strided convolutional layers for downsampling and transpose convolutional layers for up sampling. The model was trained using Mean Squared Error (MSE) loss and ADAM optimiser for 5,000 epochs, with tuned hyperparameters for the learning rate of $10^{-4}$ and batch size 32, and using the same patch-wise data augmentation strategy employed in MRF-IDDPM training.

\subsubsection{SCQ} In contrast to the previous methods, the SCQ technique \cite{fang2019supervisedmrf} takes the full length of the accelerated scans ($l=200$) as input and directly outputs the parameter maps of interest. Instead of using SVD, SCQ firstly compresses the TSMIs from $l$ to 46 with a fully connected neural network. This is then processed by two U-Nets to estimate the T1 and T2 tissue parameters, respectively. To implement the SCQ we followed the description provided in \cite{fang2019supervisedmrf}. Similar to our approach, SCQ models were trained on patches of size $64 \times 64$, while the learning rate, loss function and number of epochs were grid-searched for optimal performance. We found that a learning rate of 0.01, with MSE optimisation over 5,000 epochs, yields the best results for our data. 

\subsection{Evaluation metrics}
\label{subsec:exps_evalmetrics}
\begin{figure*}[tbh]
  \setlength{\tabcolsep}{1pt}
  \centering
  \begin{tabular}{rr}
    \footnotesize \textbf{LRTV} \hspace{0.75cm} \textbf{SCQ} \hspace{0.65cm} \textbf{DRUNet} \hspace{0.1cm} \textbf{MRF-IDDPM}  \hspace{0.05cm} \textbf{Reference} \hspace{0.15cm} 
    & \footnotesize \textbf{LRTV} \hspace{0.75cm} \textbf{SCQ} \hspace{0.7cm} \textbf{DRUNet} \hspace{0.1cm} \textbf{MRF-IDDPM}  \hspace{0.05cm} \textbf{Reference   } \hspace{0.3cm}\\
    \includegraphics[height=0.185\textheight,trim={0.25cm 0.25cm 0.25cm 0.25cm},clip]{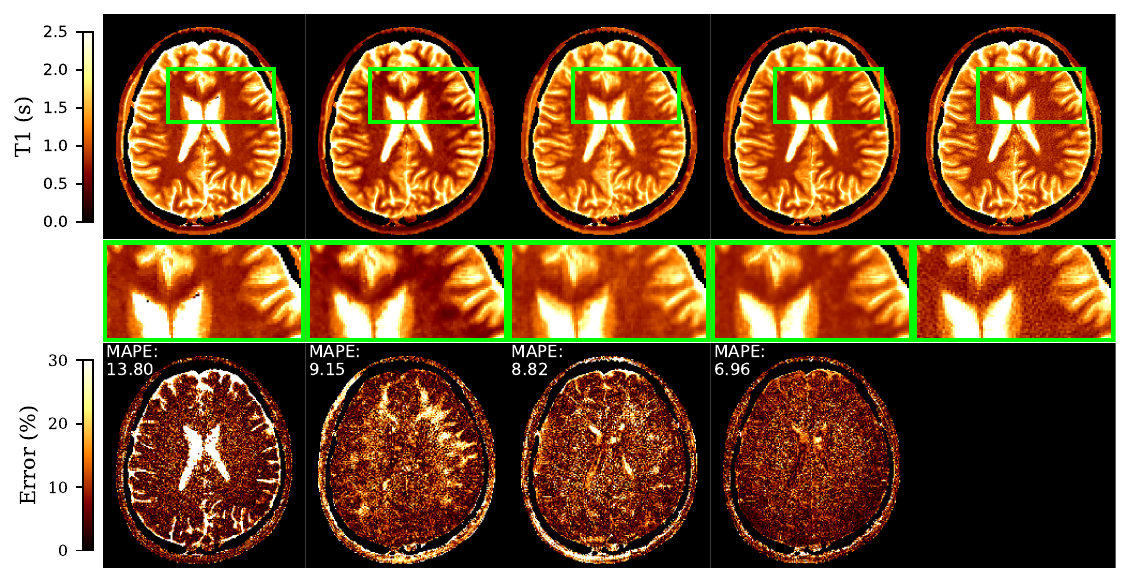}
    & \includegraphics[height=0.185\textheight,trim={0.25cm 0.25cm 0.25cm 0.25cm},clip]{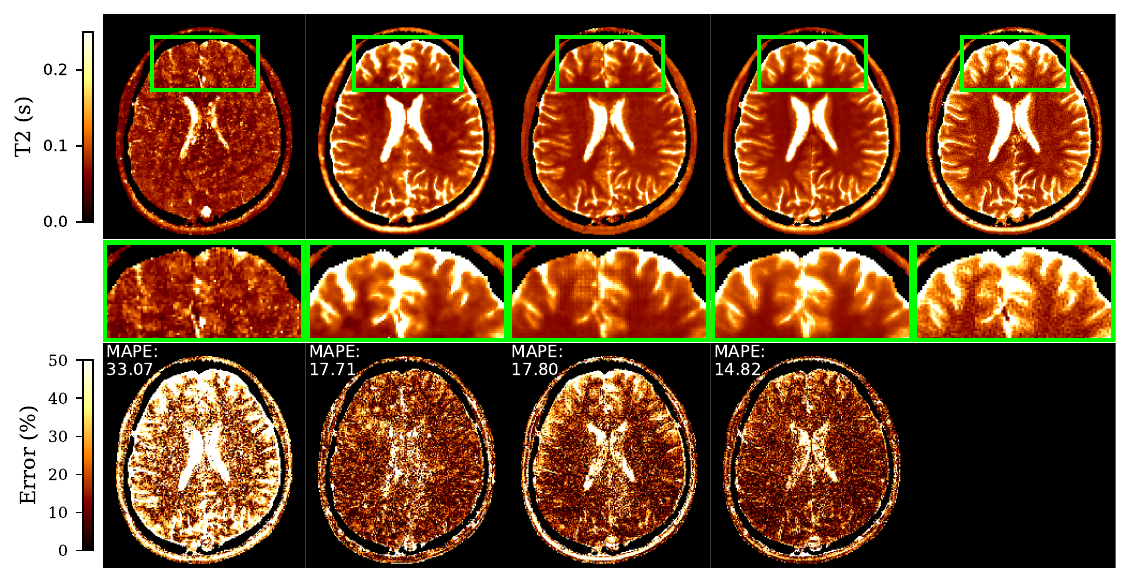} \\
    \includegraphics[height=0.185\textheight,trim={0.25cm 0.25cm 0.25cm 0.25cm},clip]{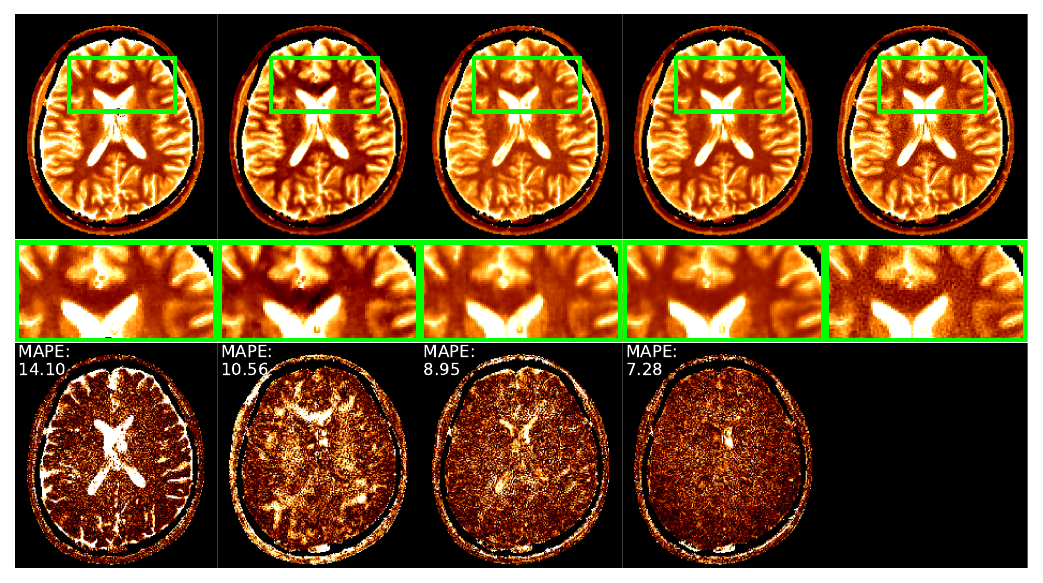}
    & \includegraphics[height=0.185\textheight,trim={0.25cm 0.25cm 0.25cm 0.25cm},clip]{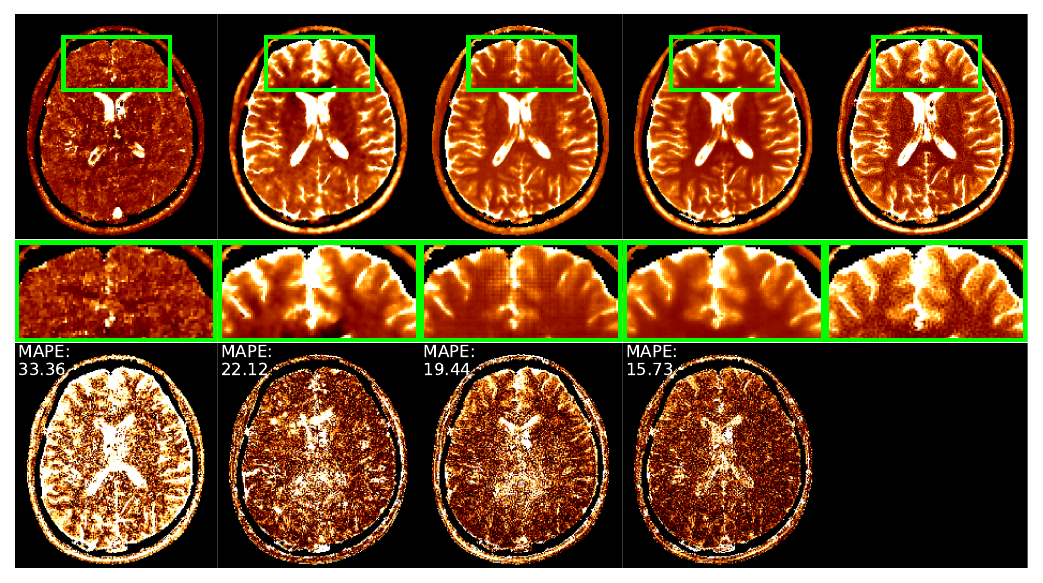} \\
    \includegraphics[height=0.185\textheight,trim={0.25cm 0.25cm 0.25cm 0.25cm},clip]{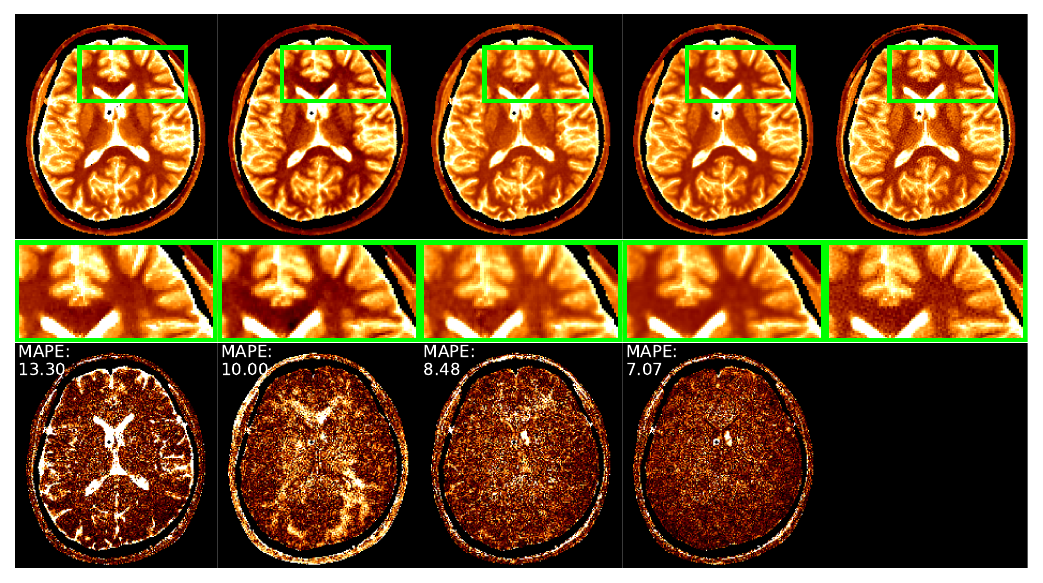}
    & \includegraphics[height=0.185\textheight,trim={0.25cm 0.25cm 0.25cm 0.25cm},clip]{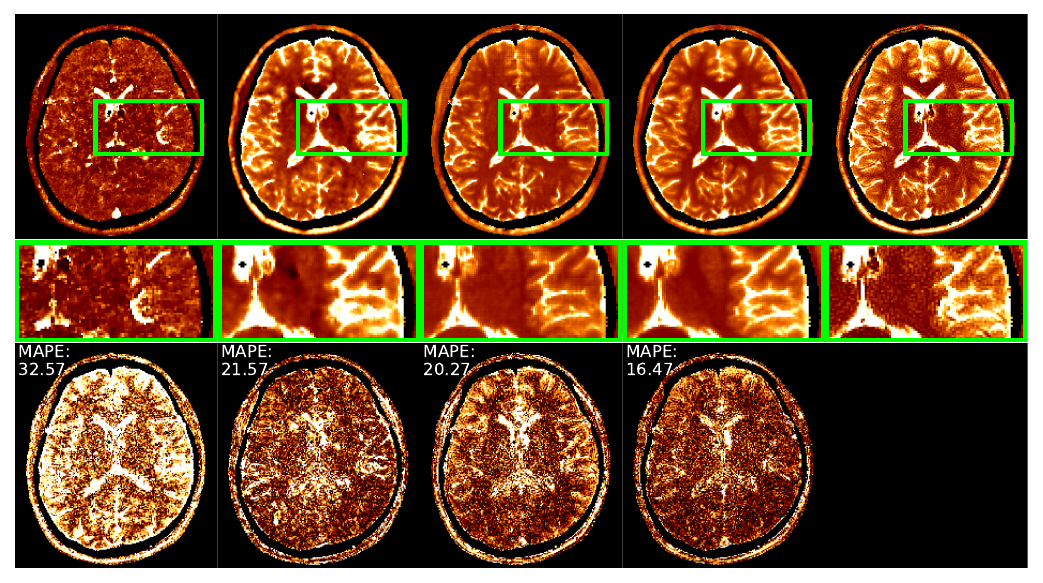} \\
  \end{tabular}
    \caption{
    Reconstructed T1 (left panel) and T2 (right panel) maps using different methods (columns) for three representative brain slices from the evaluation set. Every three rows correspond to one brain slice, displaying (from top to bottom): the full reconstructed map, a zoomed-in view of a region of interest (green box), and the corresponding percentage error map. Electronic zooming is recommended for detailed inspection.
    }
    \label{fig:tissue_maps}
\end{figure*}

\begin{figure}[t]
    \centering
    \includegraphics[width=0.48\textwidth,trim={0.3cm 0.25cm 0.25cm 0.25cm},clip]{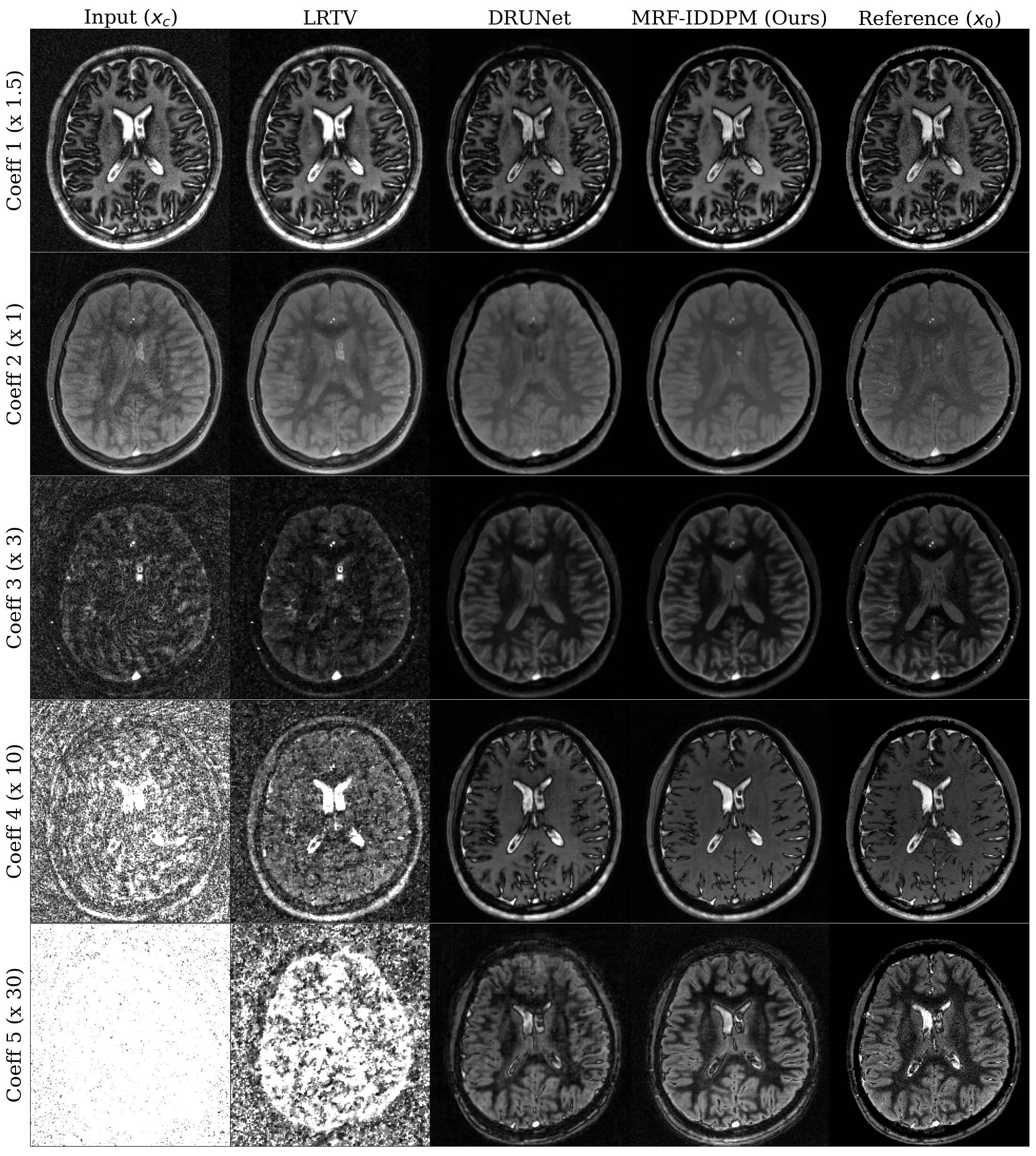} 
    \caption{Magnitudes of the reference and reconstructed TSMIs (SVD compressed to $s=5$) from one representative slice in the evaluation dataset, using different methods. Methods aim to restore the corrupted input (condition image) from a gridding reconstruction.}
    \label{fig:tsmis}
\end{figure}

We evaluate the performance of tested algorithms for T1 and T2 maps, and the TSMI reconstructions. TSMI metrics are not available for SCQ, as this method directly outputs parameter maps. For evaluations, we used two types of masks: i) head masks, including brain and surrounding non-brain/skull tissues (excluding background) for qualitative demonstrations in Figures~\ref{fig:tissue_maps}, \ref{fig:tsmis} and~\ref{fig:uncertainty_t1_t2}, and ii) skull-stripped (brain-only) masks for deriving quantitative metrics in Table~\ref{tab:metrics_selected_slices}. Some reconstruction metrics rely on error maps $\boldsymbol\varepsilon \in \mathbb{R}^{n\times s}, {\boldsymbol\varepsilon_{v,i}}={\boldsymbol\Phi}_{v,i} - \hat{\boldsymbol\Phi}_{v,i}$ between the reference $\boldsymbol\Phi_{v,i}$ and the predicted $\hat{\boldsymbol\Phi}_{v,i}$ values for the voxel $v$ and channel $i$. Note that channel $s=5$ for TSMI reconstructions, while $s=1$ for each parameter map. For clarity, we drop the index $i$ in the T1/T2 metrics. To evaluate T1 and T2 mapping performances we report the Mean Absolute Percentage Errors:

\begin{equation}
    \text{MAPE} = \frac{100}{M}\sum_{v\in \mathcal{M}}\left|\frac{{\boldsymbol\varepsilon_{v}}}{{\boldsymbol\Phi}_{v}}\right|,
\end{equation} 

and Root Mean Square Errors $\text{RMSE} =\sqrt{\frac{1}{M}\sum_{v\in \mathcal{M}}{|\boldsymbol\varepsilon}_{j}|^2}$, where $\mathcal{M}$ represents the spatial mask and $M={\text{size}(\mathcal{M})}$. We also report the Normalized Root Mean Squared Errors $\text{NRMSE} = \frac{1}{s}\sum_{i=1}^{s}{\|{\boldsymbol\varepsilon}_{v,i\in \mathcal{M}}\|_{2}}/{\|{\boldsymbol\Phi}_{v,i\in \mathcal{M}}\|_2}$, and the Mean Structural Similarity Index (SSIM) for T1 and T2 mappings, and the TSMI reconstructions. To deal with the multi-channel complex nature of the TSMI data, we compute the SSIM for the imaginary and real components for each channel separately, and report their averages.

\subsection{Results and Discussions}
\label{subsec:exps_results_discussions}
Table~\ref{tab:metrics_selected_slices} quantitatively compares the performance of MRF-IDDPM against other reconstruction baselines. To complement the quantitative results, Figure~\ref{fig:tissue_maps} presents the reconstructed T1 and T2 maps from MRF-IDDPM alongside baseline methods (SVDMRF is excluded due to its less competitive performance). Figure~\ref{fig:tsmis} illustrates the reconstructed TSMI across $s=5$ channels comparing MRF-IDDPM results to reference images and baseline methods (with SCQ excluded, as it does not reconstruct TSMI). 

Both quantitative and qualitative comparisons demonstrate the superiority of deep learning methods over SVDMRF and LRTV, due to the effective use of data-driven priors for reconstruction. While DRUNet performs competitively against SCQ, both methods fall short compared to the diffusion-based reconstruction. The proposed MRF-IDDPM consistently outperforms the baselines across all tested metrics for T1, T2, and TSMI reconstruction (Table~\ref{tab:metrics_selected_slices}). It also achieves the best qualitative trade-off between image sharpness and artifact removal (Fig.~\ref{fig:tissue_maps} and Fig.~\ref{fig:tsmis}). We hypothesize that a key factor contributing to the strong performance of MRF-IDDPM may be the inclusion of attention blocks. 
Attention mechanisms are known to enhance image reconstruction performance by modelling longer-range pixel dependencies more effectively than traditional CNNs~\cite{ozdenizci2023patchddpm, vaswani2017attention}.

Visual inspection of the reconstructed maps (Figure~\ref{fig:tissue_maps}) indicates seemingly good performance for T1 estimations, with errors becoming more apparent with the aid of the error maps. However artifacts are much more noticeable in the T2 reconstructions. Despite their sharpness, the reference T2 maps exhibit some high-frequency artifacts, as also observed in the reference TSMIs in Figure~\ref{fig:tsmis}. This occurs because the MRF data ($l=1000$) used for generating the reference TSMIs and Q-Maps was also undersampled across the k-space and for reference image reconstruction we applied a very mild Total Variation penalty to avoid over-smoothing the anatomies, thereby balancing sharpness and artefacts. In contrast, IDDPM effectively removes artifacts while incurring a smaller trade-off in sharpness, even with 5x accelerated scans. Note that the LRTV or other baselines underperform MRF-IDDPM in this acceleration regime. A similar pattern is observed in the TSMI reconstructions (Figure~\ref{fig:tsmis}), where the last two channels, arguably the weakest and most difficult to restore, are better reconstructed by MRF-IDDPM than the baselines. Overall, the MRF-IDDPM implementations provide the cleanest reconstructions among all techniques.

\subsection{Ablation Studies}
\label{subsec:exps_ablation}
Diffusion models offer various configuration choices for both training and sampling, with their optimal selection highly dependent on the specific application and task. In the following sections, we outline the key parameters that were employed in our experiments. The baseline setup, as described in Section~\ref{subsec:training}, serves as the reference for the following ablation experiments, where we assess the effect of altering individual parameters from the base configuration. Metrics report the average performance on test dataset. Unless otherwise stated, here samples were reconstructed from a single realization (i.e. the reverse process is run for only one $\x_T$).

\begin{table}[t!]
    \centering
    \scriptsize 
    \begin{tabular}{c|cc|cc|ccc}
    \hline
    \hline
         \textbf{Patch} & \multicolumn{2}{ c|}{\textbf{Time (s) $\downarrow$}} & \multicolumn{2}{ c|}{\textbf{MAPE (\%)} $\downarrow$} & \multicolumn{3}{ c}{\textbf{NRMSE} $\downarrow$} \\ \cline{2-8}
            \textbf{Size} & \textbf{T} & \textbf{I} & \textbf{T1} & \textbf{T2} & \textbf{T1} & \textbf{T2} & \textbf{TSMI} \\ \hline 
         32      &   \textbf{371} &  \textbf{35} & 8.78 & 19.31 & 0.1389 & 0.4329 & 0.6165  \\
         64      &   642 &  95 & 7.19 & 17.64 & 0.1000 & 0.3669 & 0.4142 \\ 
         128     & 2,210 & 147 & \textbf{6.19} &  \textbf{16.44} &  \textbf{0.0875} &  \textbf{0.3238} & \textbf{0.3201} \\ \hline \hline
    \end{tabular}
    \caption{The runtimes and the T1, T2 and TSMI reconstruction errors of MRF-IDDPM models trained on different spatial patch dimensions. Training time (T) is to complete 1K iterations, inference time (I) is to reconstruct one image. Metrics are averaged over test dataset.}
    \label{tab:metrics_image_size}
\end{table}

\subsubsection{Effect of patch size}
\label{subsubsec:ablation_patchsize}
A key aspect of our approach is the use of patch-based training and sampling. In this study, we investigated the impact of different patch sizes on the performance of our method. Three input sizes were considered: $p=32 \times 32$, $64 \times 64$ and $128 \times 128$. Accordingly, three MRF-IDDPM models were trained, all following the same specifications as in Section~\ref{subsec:training}. The results of this study are summarised in Table~\ref{tab:metrics_image_size}. We observe that models trained on larger patch sizes can produce more accurate reconstructions, as they capture long-range pixel dependencies across broader spatial regions. However, training and sampling for these models require significantly more time, making them less ideal. In this study, we found that using $64 \times 64$ patches provides a favourable balance between efficient training time and high reconstruction quality, with a difference of just $\sim 1\%$ in the T1/T2 MAPE compared to the model trained with a patch size of $128 \times 128$.

\subsubsection{Effect of patch stride size}
\label{subsubsec:ablation_patch_stride}
\begin{figure}[t!]
    \centering
    \begin{subfigure}{.5\textwidth}
        \centering
        \includegraphics[width=0.5\linewidth,trim={0 0 0.2cm 0},clip]{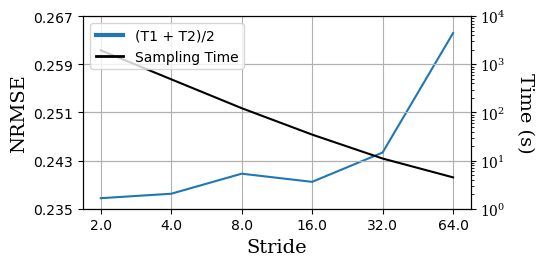}  
        \includegraphics[width=0.45\linewidth]{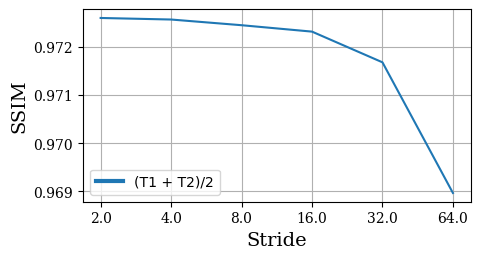}  
    \end{subfigure}
    \begin{subfigure}{.5\textwidth}
        \centering
        \includegraphics[width=0.45\linewidth]{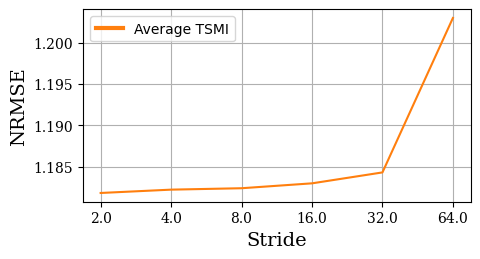} 
        \includegraphics[width=0.45\linewidth]{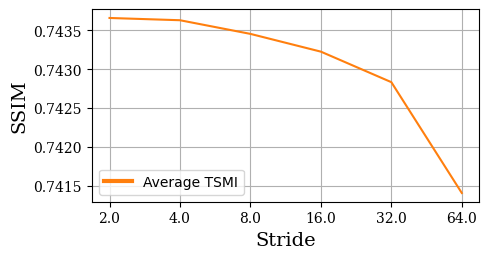}  
    \end{subfigure}
    \caption{Effect of stride size on the reconstruction time (black curve), as well as the reconstruction NMRSE (left) and SSIM (right) metrics for TSMI (bottom) and averaged T1 and T2 maps (top, labelled as (T1+T2)/2). Metrics are averaged over the test dataset.}
    \label{fig:stride_effect}
\end{figure}

In our approach, the final output is assembled from individual reconstructed patches, with quality determined by both patch size and stride. Selecting the degree of overlap, i.e., the patch stride, requires balancing image quality with reconstruction time. High overlap improves reconstruction quality but can drastically increase the runtime. Conversely, using low to no overlap may speed up sampling but often leads to blocking image artifacts and degraded reconstruction performance. Therefore, an optimal stride size is necessary to achieve high-quality outputs while maintaining computational efficiency. We explored this trade-off by varying the stride size during IDDPM image reconstruction from 2 to 64 (non-overlapping). Figure~\ref{fig:stride_effect} displays the reconstruction times alongside the performance metrics for T1, T2, and TSMI for each tested stride size. Both MAPE and SSIM metrics indicate a good trade-off between reconstruction accuracy vs. computation time for a stride size of 8. Stride sizes larger than 8 made reconstruction quality decline. Smaller stride sizes did not improve the reconstruction much, while increasing the computation time  more than one order of magnitude.

\begin{figure}[t!]
    \centering
    \begin{subfigure}{.5\textwidth}
        \centering
        \includegraphics[width=0.5\linewidth]{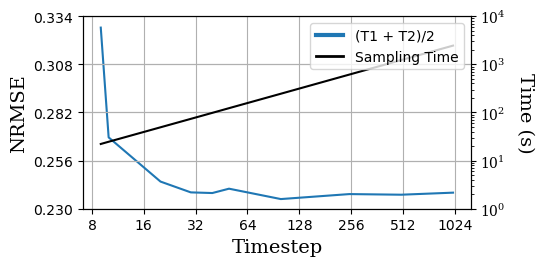}  
        \includegraphics[width=0.45\linewidth]{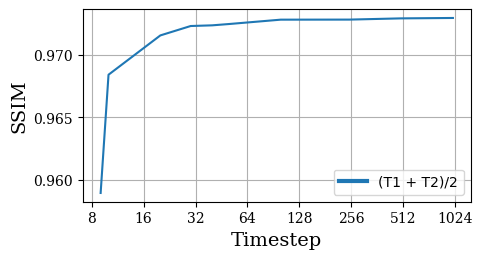}  
    \end{subfigure}
    \begin{subfigure}{.5\textwidth}
        \centering
        \includegraphics[width=0.45\linewidth]{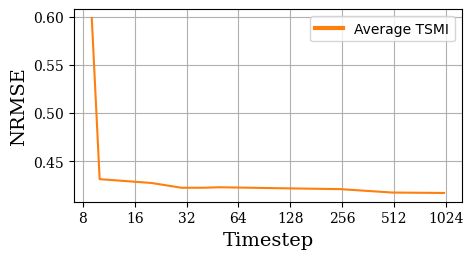} 
        \includegraphics[width=0.45\linewidth]{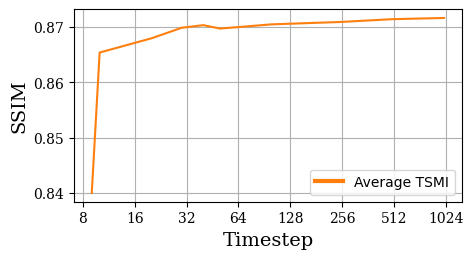}  
    \end{subfigure}
    \caption{Effect of timesteps (K) on the reconstruction time (black curve), as well as the reconstruction NMRSE (left) and SSIM (right) metrics for TSMI (bottom) and averaged T1 and T2 maps (top). Metrics are averaged over the test dataset.}
    \label{fig:timestep_effect}
\end{figure}

\begin{figure*}[t!]
    \centering
    \begin{subfigure}[t]{0.49\textwidth}
        \centering
        \includegraphics[width=0.985\textwidth,trim={0.25cm 0.25cm 0 0.25cm},clip]{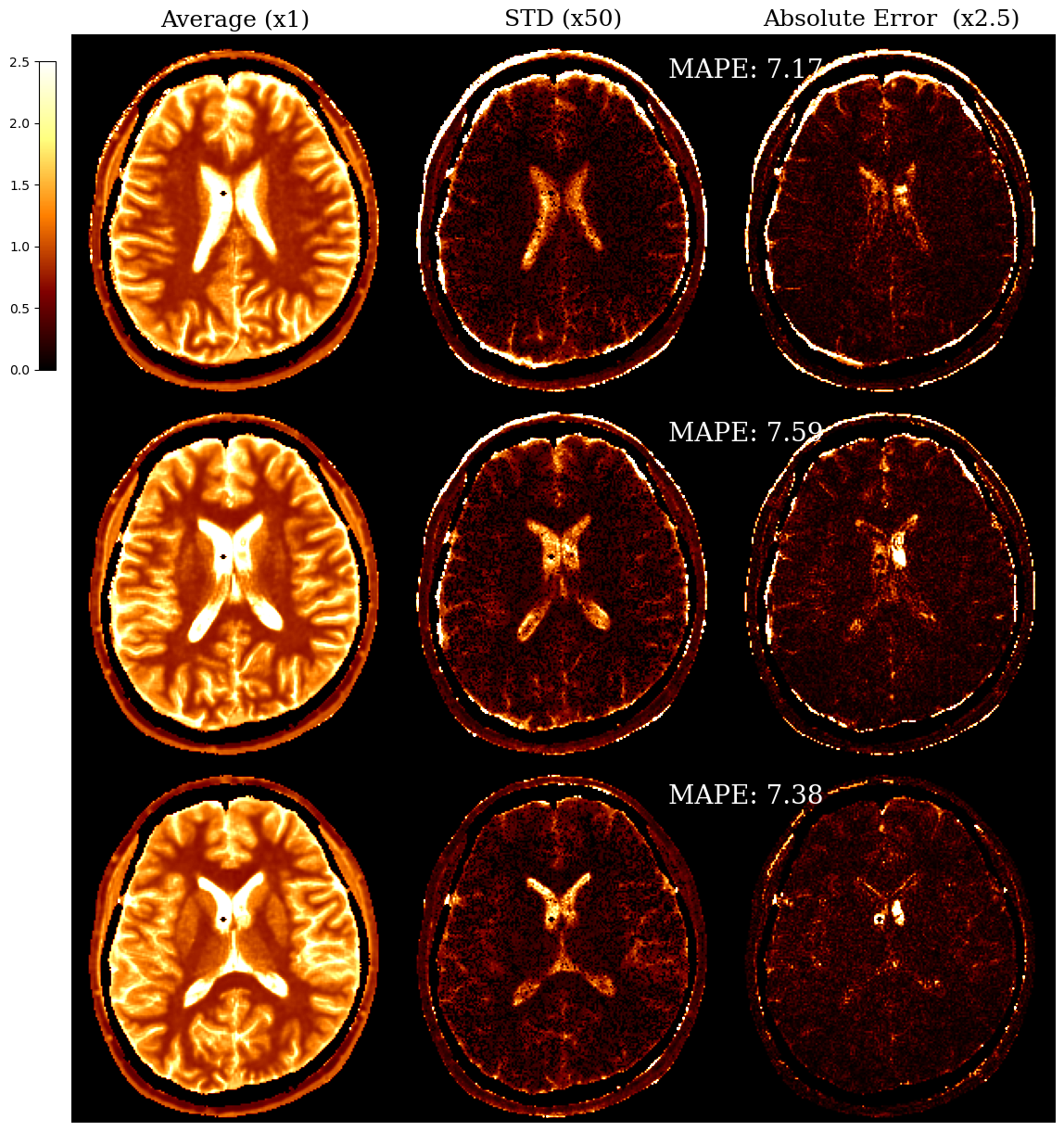}
    \end{subfigure}
    \begin{subfigure}[t]{0.49\textwidth}
        \centering
        \includegraphics[width=\textwidth,trim={0 0.25cm 0.25cm 0.25cm},clip]{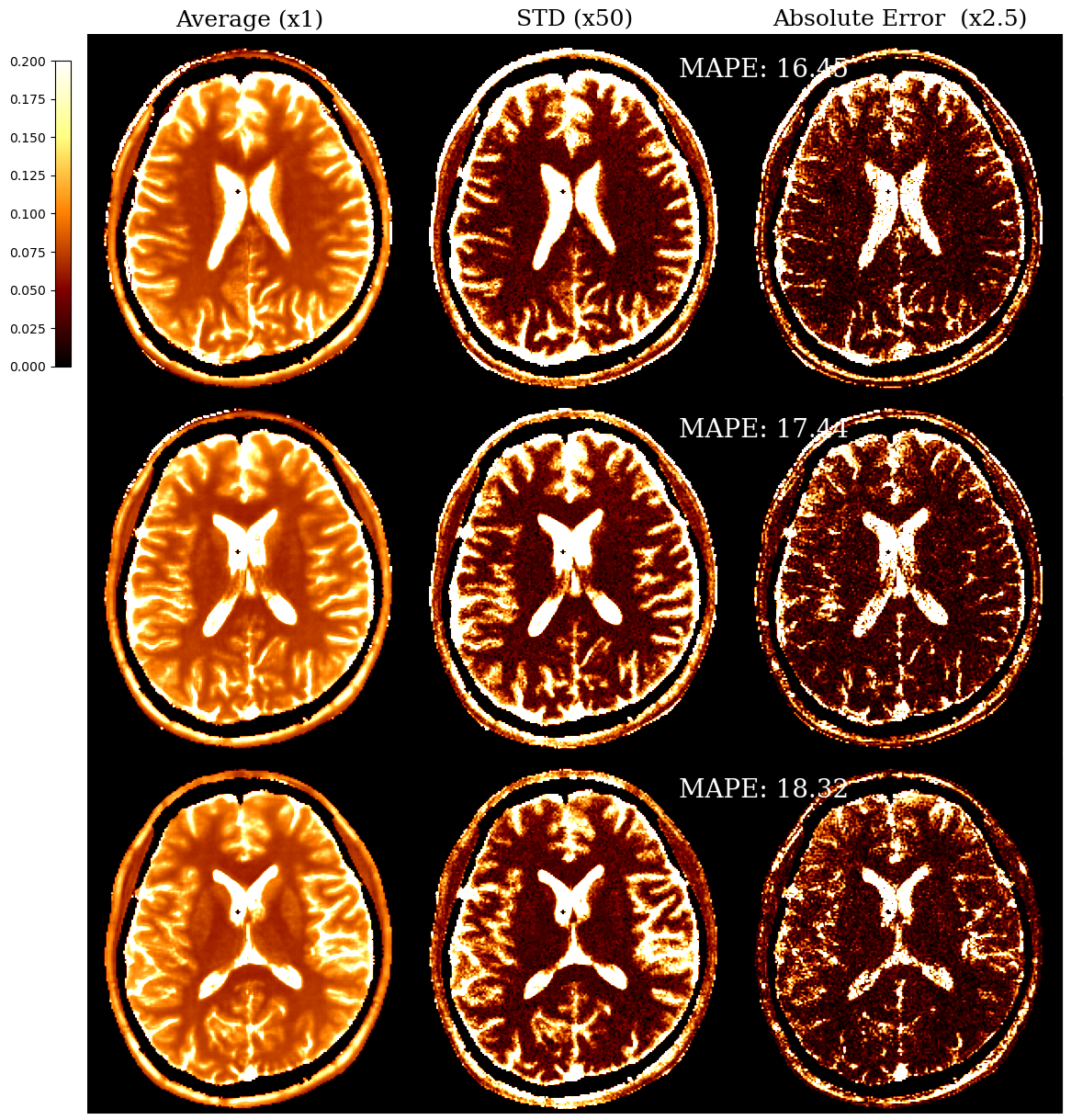}
    \end{subfigure}
    \caption{Average, standard deviation (STD), and absolute error maps for T1 (left panel) and T2 (right panel) reconstructions of three representative brain slices in the evaluation set. Each row corresponds to a different brain slice. The MRF-IDDPM algorithm was used to generate 10 reconstructions per slice, each initiated with a different random noise $\x_T$. Average and STD maps show the pixel-wise mean and standard deviation across these 10 reconstructions. Absolute error maps display the absolute difference between the average reconstruction and the ground truth reference from Figure~\ref{fig:tissue_maps}.
    }
    \label{fig:uncertainty_t1_t2}
\end{figure*}

\subsubsection{Number of time steps}
\label{subsubsec:sampling_backbone}
We also examined the number of IDDPM time steps $K$ required for sampling accurate MRF reconstructions. Results are presented in Figure~\ref{fig:timestep_effect}, showing NRMSE and SSIM as a function of $K$ for the estimations of T1/T2 maps and the TSMIs. As expected, the time required for a complete pass of the reverse process for a single image increases linearly with $K$. Reconstruction NRMSEs drop rapidly with only a few sampling steps, plateauing promptly after this fall. Similarly, the SSIM metrics improve rapidly, although they plateau later than the NRMSE. The plots suggest that 30 sampling steps are enough for the reverse process. We used $K=50$ in all other experiments throughout, as a common choice in the literature~\cite{ozdenizci2023patchddpm,huang2024padmr} while still achieving a good trade-off between the reconstruction accuracy and runtime.

\subsubsection{Estimation of missing time frames}
\label{subsubsec:ablation_gts}
So far, our base models assume that the input $\x_c$ and target $\x_0$ TSMIs have the same truncated time length ($l=200$). To expand on this, we also trained models (with the same input) to reconstruct outputs $\hat{\x}_0$ of length $l=1000$, matching the longer reference acquisition sequence. These two configurations are referred to as Task 1 ($l=200$) and Task 2 ($l=1000$). Outputs of both tasks used SVD dimensionality reduction but with different bases, $\textbf{V}_{200}$ and $\textbf{V}_{1000}$, corresponding to dictionaries with truncated and untruncated fingerprints, respectively. For each task, we trained both MRF-IDDPM and DRUNet models and compared their test performances in Table~\ref{tab:metrics_different_targets}. 

\begin{table}[htb]
    \centering
    \scriptsize 
    \begin{tabular}{c|c|cc|ccc}
    \hline \hline
         \multirow{ 2}{*}{\textbf{Method}} & \multirow{ 2}{*}{\textbf{Task}} & \multicolumn{2}{ c|}{\textbf{MAPE (\%) $\downarrow$}} & \multicolumn{3}{ c}{\textbf{NRMSE} $\downarrow$} \\ \cline{3-7}
            & & \textbf{T1} & \textbf{T2} & \textbf{T1} & \textbf{T2} & \textbf{TSMI} \\ \hline
         \multirow{ 2}{*}{\textbf{DRUNet} \cite{zhang2021drunet}}
            & 1 &  8.93 & 21.07 & 0.1285 & 0.4458 & 0.5633 \\ 
            & 2 & 13.98 & 24.71 & 0.1651 & 0.4473 & 0.5907 \\ \hline 
         \multirow{ 2}{*}{\textbf{Ours}} 
            & 1  & 7.19 & 17.64 & 0.1000 & 0.3669 & 0.4142 \\ 
            & \textbf{2}  &  \textbf{6.46} & \textbf{16.94} & \textbf{0.0892} & \textbf{0.3167} & \textbf{0.3449} \\ \hline \hline
    \end{tabular}
    \caption{Reconstruction errors for TSMI and T1/T2 parameter maps using MRF-IDDPM and DRUNet models trained on two tasks:  1) TSMI restoration only and 2) TSMI restoration and mapping to a longer timeframe.}
    \label{tab:metrics_different_targets}
\end{table}

The MRF-IDDPM demonstrates a marginal improvement in reconstruction metrics for Task 2, with gains of $\sim 1\%$ in T1/T2 MAPE and $\sim 7\% $ in TSMI NRMSE. This could be because longer time series give more discriminant signals for different T1/T2 values and so reconstructions were able to benefit from this. However, this advantage was not observed in the standard CNN implementation. DRUNet performed better in reconstructing TSMIs at $l=200$, but its performance declined  on Task 2. Notably, MRF-IDDPM outperformed  DRUNet across both tasks.

\subsubsection{Uncertainty Maps}
\label{subsubsec:uncertainty}
Conditional diffusion models are shown to be useful to additionally produce a notion of uncertainty map alongside their reconstructed images (for example, see~\cite{luo2023bayesian_uncertainty,liu2023dolce}). This uncertainty can be visualized as a variance map of the output images obtained from multiple runs/samples of the reconstruction process in Algorithm~\ref{alg:inference}. Each run starts with a different random draw of the initial noise $\x_T \sim \mathcal{N}(\textbf{0},\textbf{Id})$, but all use the same conditioning image $\x_c$. This allows for the inspection of the model by assessing the variations across the reconstructed samples. To show this, we ran the reverse diffusion process to reconstruct 10 TSMIs (and the corresponding T1 and T2 maps), using the same low-quality conditioning image but starting with different initial random noise $\x_T$ each time. We then computed the pixel-wise standard deviation across the 10 sampled outputs and compared it to the absolute error between the averaged reconstructions and the reference. Figure~\ref{fig:uncertainty_t1_t2} shows the averaged T1 and T2 reconstructions, along with the variance and absolute error maps for three selected brain slices in Figure~\ref{fig:tissue_maps}. The figures demonstrate clear correlations between the variance and error maps, with higher errors occurring in regions of higher variance. Large errors are primarily occurring in the cerebrospinal fluid and surrounding fat/skull areas, and less so in white and gray matter. 

\subsection{Limitations and Future Work}
\label{subsec:exps_limitations_future}
Our study was restricted to a small in-house dataset of healthy subjects, and thus generalization to other datasets is yet to be examined, including data containing pathologies, brain diseases, or different anatomical regions than the brain. Furthermore, our approach was tested on 2D imaging data and considering future extensions to 3D MRF acquisitions e.g.~\cite{gomez2019qtimrf,cao2022tgas,liao20173dmrf} would require accounting for spatial context and correlations across 3D volumes, as well as increased computational resources. Processing images at the patch level as proposed, or using latent space approaches like encoder-decoder models could help manage the high dimensionality and memory demands for 3D reconstructions. Lastly, in our current pipeline, Bloch response constraints were applied during the parameter quantification stage, after TSMI reconstruction. While achieving good performance, further improvements might be achieved by incorporating these constraints, along with k-space consistency, during the TSMI reconstruction process itself (see e.g.~\cite{hamilton2022dipmrf,li2023learnedtensor,zhang2024large3d} for non-DDPM based MRF models that account for these constraints, also~\cite{gungor2023adaptive,cui2024spirit} for MRI reconstructions DDPM models that account for k-space consistency).

\section{Conclusions}
\label{sec:conclusions}
This paper introduced a novel IDDPM-based model for the enhanced reconstruction of parameter maps from fast transient-state qMRI techniques, such as MRF. We evaluated our method on in-vivo brain scans from healthy volunteers, where the MRF acquisition sequence was retrospectively truncated by a factor of 5, i.e. 5-fold acceleration. Both qualitative and quantitative results demonstrated that the proposed approach, MRF-IDDPM, outperforms established deep learning and conventional compressed sensing algorithms for MRF reconstruction. The MRF-IDDPM's probabilistic framework allows for sampling variations across reconstructed images, which can be useful for generating uncertainty maps
and to assess confidence in the reconstruction at each pixel. To reduce computational overhead, we employed an efficient DDPM sampling strategy and utilized patch-level image processing during both training and reconstruction, allowing these steps to be performed on modest GPUs within a reasonable runtime.





%
%
%
\bibliographystyle{IEEEtran}
\bibliography{references}


\end{document}